# Forecasting Short-term load using Econometrics time series model with T-student Distribution

Kasun Chandrarathna*[1], Arman Edalati [2], AhmadReza Fourozan tabar[3]

*[1]Student, Department of Electrical Engineering, Texas Tech University, Lubbock, TX, USA

[2]Student, Industrial Engineering Department, Damghan University, Damghan, Iran

[3]Asisstant professor, Electrical Engineering Department, Marvdasht Azad University, Marvdasht, Iran

## ABSTRACT

By significant improvements in modern electrical systems, planning for unit commitment and power dispatching of them are two big concerns between the researchers. Short-term load forecasting plays a significant role in planning and dispatching them. In recent years, numerous works have been done on Short-term load forecasting. Having an accurate model for predicting the load can be beneficial for optimizing the electrical sources and protecting energy. Several models such as Artificial Intelligence and Statistics model have been used to improve the accuracy of load forecasting. Among the statistics models, time series models show a great performance. In this paper, an Autoregressive integrated moving average (SARIMA) - generalized autoregressive conditional heteroskedasticity (GARCH) model as a powerful tool for modeling the conditional mean and volatility of time series with the T-student Distribution is used to forecast electric load in short period of time. The attained model is compared with the ARIMA model with Normal Distribution. Finally, the effectiveness of the proposed approach is validated by applying real electric load data from the Electric Reliability Council of Texas (ERCOT).

## I. INTRODUCTION

Money talks is a well-known expression that can be applied to the cost of electricity generation. Over-generation of electricity can cause a huge waste of energy and money. While under-generation can be even more damaging to the costumers and impose a huge power outage to the system and cause lots of avoidable expenses. Therefore, having precise forecasting of the load in the future like several hours or one day ahead is valuable. By having an acceptable approximation of the amount of load, the production of electrical power plants can be optimized, and the waste of energy and power outages can be minimized to acceptable ranges. Numerous works have been done on the Short-term load forecasting. It can be effective for the all the components of electrical systems such as alternators or vehicles prices [1-4]. The researchers have tried to predict the load for the short term (between 1 hour to one day ahead). Several different approaches such as machine-learning algorithms such as artificial neural networks, Support Vector Machines, and time series analysis (including Autoregressive Integrated Moving Average (ARIMA) and the Autoregressive Moving Average (ARMA)) have been applied for load forecasting, the mentioned method has been used related to electric vehicles [5-7]. In [8-9] Support Vector Machines Regression (SVR) for forecasting the load has been considered. Artificial Neural Networks (ANN) are other methods that have been used for forecasting the load and the response was compared with the classical ANN [9]. Gaussian process (GP) is another model that has been investigated in several works [10-12]. A classical ARIMA model which is a statistical time-series model was applied to the data for having an acceptable forecasted range of load [13-14].

This research aims to forecast the electrical load based on the real data which is taken from the Electric Reliability Council of Texas (ERCOT). In this paper, an Autoregressive integrated moving average (SARIMA) - generalized autoregressive conditional heteroskedasticity (GARCH) model as a powerful tool for modeling the conditional mean and volatility of time series with the T-student Distribution is used to forecast electric load for one day ahead (24 hours). The attained model is compared with the ARIMA model with Normal Distribution. Finally, the effectiveness of the proposed approach is validated by applying real electric load data from the ERCOT.

## II. METHODOLOGY

In this section, we shortly describe the method used in this study. We use real electricity load data from the Electric Reliability Council of Texas (ERCOT) obtained from the official website from 1:00 am 01/02/2011 to 11:00 pm 12/21/2018. Figure 1 displays the electricity load time-series process for the data set studied in this research.

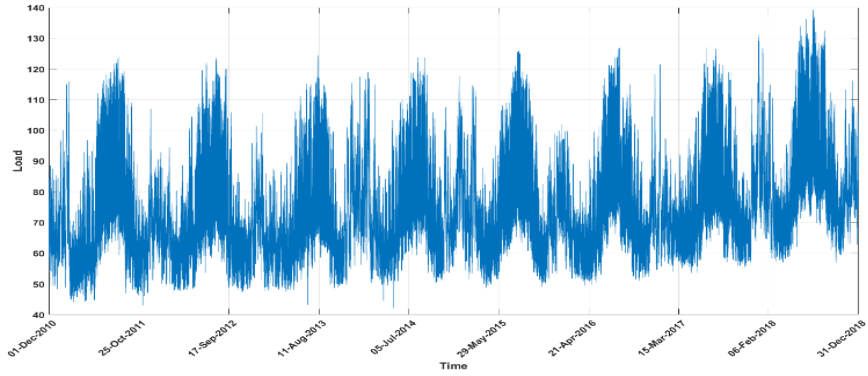

**Figure 1** Load time series process from 01/02/2011 to 12/31/2018.

To examine whether the electricity load process is non-stationary, we apply the traditional unit-root non-stationary tests. The p-value (~ 0.742) of the Augmented Dickey-Fuller test presents clear proof of the unit-root non-stationary process. In other words, the analysis confirms that load data are I(1), or integrated of order one. As the load process confers non-stationary behavior, we require to convert the load process to a stationary process using a functional form. reason behind this transformation is that we need to work with stable data to have accurate forecasting. The most important of the functional forms that we use in time series regressions is the natural logarithm. We obtain the log-returns of the load process by the logarithmic return as follows

$$r(t) = Ln \frac{S(t)}{S(t-1)} \qquad (1)$$

where S(t) is the load value at day t. Figure (2) shows time series return process.

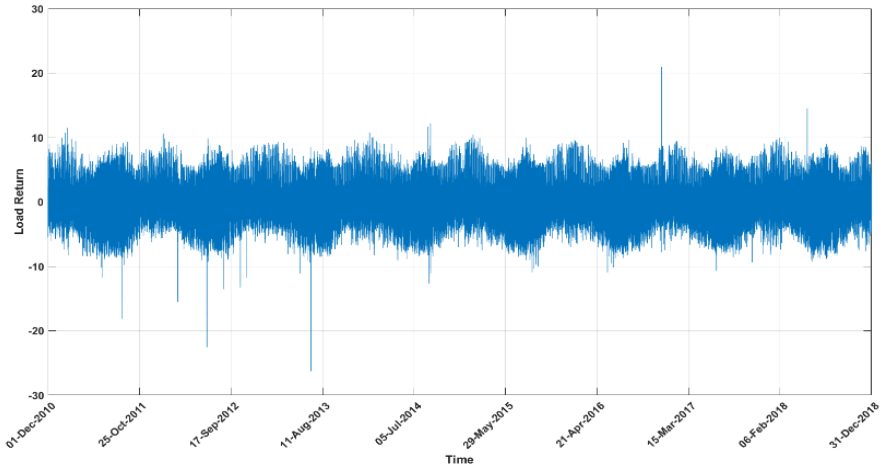

**Figure 2** Return process from 01/02/2011 to 12/31/2018

Having the transformed data (load return process), we perform the Augmented Dickey-Fuller test for checking the I(1) process. The test's p-value is approximately zero, showing a clear indication of rejecting the unit-root non-stationary process. The McLeod-Li test is performed to investigate the existence of conditional heteroscedasticity in load time series data. The test's p-value is about zero indicating the presence of conditional heteroscedasticity in the return process. Here, we should filter out the data set from serial dependence and volatility clustering. The linear and non-linear dependence in the data set can be removed by using econometrics time series models for obtaining sample innovations. In this case, we study their sample innovations obtained from the time series models instead of considering the load return process. This method has been widely used in the stock market and commodities to study how the price process behaves with other securities. To study time-varying dependence in the stock market and commodities, [15] and [16] applied sample innovations instead of the return process. The reason behind using sample innovations instead of the return time series is that we have independent and identically distributed (iid) standardized residuals in the model and presume to have an accurate prediction. Econometrics is an active field that widely used for modeling and predicting time series data. It has been widely used in engineering, economics, business, finance, statistics, physics, and applied mathematics. In finance and economics, researcher used the time series model to model and analysis load process (e.g. [17] and [18] used autoregressive generalized autoregressive conditional heteroscedasticity (AR-GARCH) models for predicting stock market prices). Time series regression models are an active field that broadly applied for modeling time series data. The method has been commonly employed in statistics, physics, and applied mathematics, statistics, economics, finance, business, physics, and engineering. Recently the employment of the time series model extended to creating social indices (e.g. [19] used ARMA (1,1)-GARCH (1,1) to form an index to measure the US citizenry's level of socioeconomic content). Some researchers applied time series for missing imputation data (e.g. [20] used continuous-time autoregressive models as an alternative method for imputing missing data). To forecasting electricity load, [21-22] applied the autoregressive moving average model (ARMA) to address the issue of modeling and forecasting electricity loads. [23] showed that the ARMA model performs significantly more reliable than the one used by the California System Operator. The ARMA model is defined as follows (see [13]).

[2]

$$\phi(L)(1-L)^d r_t = \theta(L) a_t; \quad 0 \leq d \leq 1, \tag{2}$$

where L is the lag operator , $a_t$ refers to electricity shocks, d is the fractional integration parameter, $\phi$, d, and $\theta$ are the parameters of the model estimated from the data. As we observe the presence of heteroscedasticity in the return process, we believe their result is not reliable for the period that the market experiences a high volatility period. Consequently, in this research we apply ARMA-GARCH model to remove the linear and non-linear auto correlation in the model. The p-value of the McLeod-Li test indicated the presence of heteroscedasticity in the load return process. If we don't remove the heteroscedasticity from the model, the forecast result is not reliable for the high volatility period in the load return process.

In this case, we employ ARMA model mixed with the generalized autoregressive conditional heteroskedasticity (GARCH) process to eliminate the linear and non-linear autocorrelation in the volatility of the load process. We also consider the seasonal impact on the load in our model to extract the global trends and business cycles of a time series. Because electricity loads can show large deviations, we consider a skew-normal distribution for model's innovations. Thus, our proposed model for the prediction one-day ahead load process is SARIMA-GARCH with skew-normal distribution for the model's innovations. The mathematics formula for the model suggested here is (see [24]).

$$\begin{cases} \phi(1-L^s)(1-L)^d r_t = \theta(1-L)^k a_t; \quad 0 \leq d \leq 1 \\ a_t = \varepsilon_t \sigma_t, \quad \varepsilon_t \sim iid, \\ \sigma_t^2 = \gamma + \alpha a_{t-1}^2 + \beta \sigma_{t-1}^2 \end{cases} \tag{3}$$

Where d is the fractional integration parameter, L is the lag operator, and $a_t$ refers to electricity shocks, $\varepsilon_t$ are iid variables with skew-normal distribution, $\sigma_t^2$ is the conditional variance, $\phi$, s, d, $\theta$, $\gamma$, $\alpha$, and $\beta$ are the parameters of the model estimated from the data[24]. Ggeneralized the normal distribution to allow for non-zero skewness. They defined a shape parameter to capture the heavy tail and skewness of the data set not obtained by the normal distribution. Probability density function (pdf) of the skew-normal distribution with parameter is $\alpha$ given by

$$f(x) = \frac{2}{\omega\sqrt{2\pi}} \exp\left(-\frac{(x-\epsilon)^2}{2\omega^2}\right) \int_{-\infty}^{\alpha\left(\frac{\omega-\epsilon}{\omega}\right)} \frac{1}{\sqrt{2\pi}} e^{-\frac{t^2}{2}} dt \tag{4}$$

Where $\epsilon$ is location, $\omega$ is scale, and $\alpha$ is shape (see [14]).

### III. RESULTS AND DISCUSSION

Table 1 SARMA model estimated parameters.

|  | Value | Standard Error | T-Statistic | P-Value |
|---|---|---|---|---|
| Constant | 0.001 | 0.000 | 3.363 | 0.001 |
| AR {2} | 0.228 | 0.004 | 57.668 | 0.000 |
| SAR {24} | 0.345 | 0.002 | 166.120 | 0.000 |
| MA {3} | 0.029 | 0.004 | 7.278 | 0.000 |
| SMA{24} | -0.878 | 0.001 | -593.510 | 0.000 |

The estimated values for the variance model are reported in Table 2. The GARCH model parameters are significant at a 5% confidence level, as the p-values of the parameters are approximately zero.

**Table 2** GARCH model estimated parameters

|  | Value | Standard Error | T-Statistic | P-Value |
|---|---|---|---|---|
| Constant | 0.370 | 0.003 | 135.360 | 0.00 |
| GARCH{1} | 0.250 | 0.004 | 65.207 | 0.00 |
| ARCH{1} | 0.562 | 0.005 | 106.190 | 0.00 |

In this section, we implement time series models to the load return process to predict 24 hours of data. We seek for the most suitable model by analyzing via AIC and BIC. The AIC and BIC criteria point to an ARIMA(2,0,3) model Seasonally Integrated with Seasonal AR(24) and MA(24), while the variance model is GARCH(1,1). The estimated parameters of the SARIMA model with the T-statistics and P-values are summarized in Table 1 reports. The p-values of the model estimated parameters are about zero, indicating statistically significant of all estimated parameters at a 5% confidence level.

We used the data set form 01/02/2011 to 31/12/2018 for electricity loads collected from the ERCOT. The recommended SARIMA-GARCH model with the Skew-Normal Distribution is implemented to the load return data from ERCOT. We employ the SARIMA-GARCH model with the Normal Distribution to matched results with our model. For the sake of a fair comparison, the forecast for the 11/11/2018 is chosen. Figure (3) displays the waveform and the forecasted trend of one day ahead of electricity loads for ERCOT. The values for the next 24 hours of ERCOT is presented in Table 3. The real values and compared model predicated values are two other items of the forecasting process are given in Table 2.

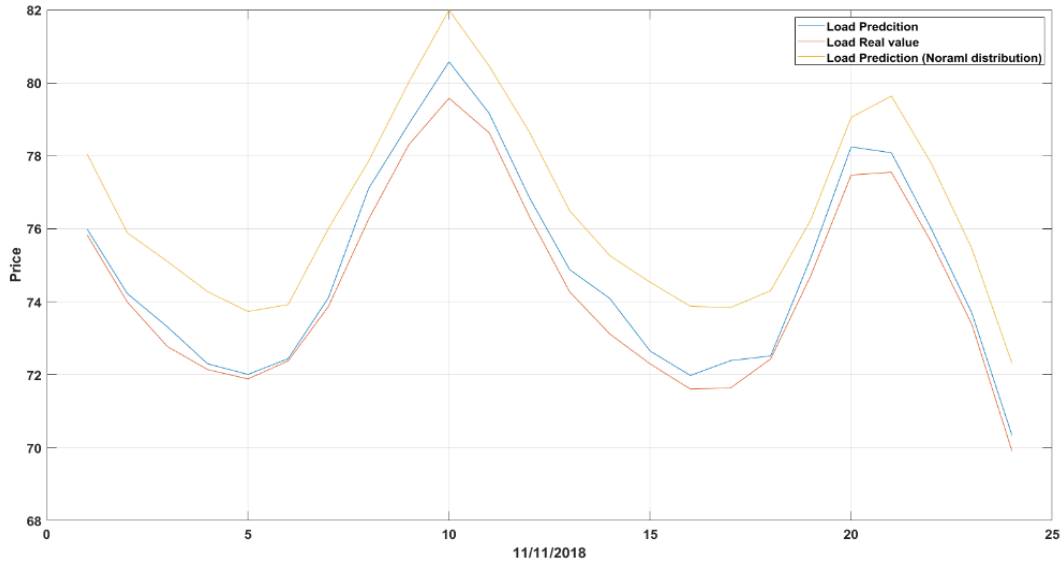

**Figure 3** Waveform and the forecasted trend of one day ahead of electricity loads for ERCOT.

As we observe from Figure1, the recommended method beats the other examined model. The square errors for the proposed model and the checked model are 0.2803 and 4.2024, respectively. The absolute errors of the suggested model (0.4583) and the reviewed model (2.0314) prove that the recommended method's precision is better than the other one. This finding demonstrates that the forecast accuracy can be significantly improved by the suggested method.

## IV. CONCLUSION

In this article, we offered a new load forecasting method based on SARIMA-GARCH models with the Skew-Normal Distribution as electricity loads exhibit large deviations. The SARIMA-GARCH method is considered for the Electric Reliability Council of Texas load prediction in the USA market. We used the most recently published loads for testing the proposed model. The results of the recommended model were compared with RIMA-GARCH with normal distribution for errors. The results clearly demonstrated that the suggested method is far more accurate than the other forecast method.

Table 3 The prediction values for the next 24 hours of ERCOT

| Date | Load | Proposed Model | Checked Model |
|---|---|---|---|
| 11/11/2018 0:00 | 75.83555 | 75.99298085 | 78.05071233 |
| 11/11/2018 1:00 | 73.99142 | 74.22428581 | 75.88984582 |
| 11/11/2018 2:00 | 72.76974 | 73.30858314 | 75.10271599 |
| 11/11/2018 3:00 | 72.14082 | 72.29787603 | 74.27351882 |
| 11/11/2018 4:00 | 71.88511 | 72.00966931 | 73.73325431 |
| 11/11/2018 5:00 | 72.37929 | 72.43900542 | 73.92281289 |
| 11/11/2018 6:00 | 73.86816 | 74.10826354 | 76.00345436 |
| 11/11/2018 7:00 | 76.26411 | 77.10171913 | 77.85683977 |
| 11/11/2018 8:00 | 78.30527 | 78.87609967 | 80.01767151 |
| 11/11/2018 9:00 | 79.57778 | 80.57674503 | 81.99401079 |
| 11/11/2018 10:00 | 78.62978 | 79.162164 | 80.45705114 |
| 11/11/2018 11:00 | 76.33346 | 76.85246984 | 78.64864993 |
| 11/11/2018 12:00 | 74.27119 | 74.88054858 | 76.49227091 |
| 11/11/2018 13:00 | 73.11548 | 74.08897677 | 75.26318849 |
| 11/11/2018 14:00 | 72.29818 | 72.64994457 | 74.5359343 |
| 11/11/2018 15:00 | 71.60746 | 71.98062829 | 73.87879707 |
| 11/11/2018 16:00 | 71.63582 | 72.38682977 | 73.83935234 |
| 11/11/2018 17:00 | 72.43516 | 72.51585023 | 74.3067652 |
| 11/11/2018 18:00 | 74.72569 | 75.21242705 | 76.25361391 |
| 11/11/2018 19:00 | 77.47444 | 78.24390751 | 79.05406518 |
| 11/11/2018 20:00 | 77.55057 | 78.08465473 | 79.63807074 |
| 11/11/2018 21:00 | 75.62043 | 75.98269319 | 77.77782385 |
| 11/11/2018 22:00 | 73.38529 | 73.69321887 | 75.46011838 |
| 11/11/2018 23:00 | 69.90736 | 70.3368659 | 72.31015298 |